\input{aipcheck}

\documentclass[
    ,final            
  ]
  {aipproc}

\layoutstyle{6x9}

\begin{document}
\newcommand{\begit}{\begin{itemize}}
\newcommand{\enit}{\end{itemize}}
\newcommand{\begen}{\begin{enumerate}}
\newcommand{\enen}{\end{enumerate}}
\newcommand{\beq}{\begin{equation}} 
\newcommand{\eeq}{\end{equation}} 
\newcommand{\beqa}{\begin{eqnarray}} 
\newcommand{\eeqa}{\end{eqnarray}} 
\def\lesssim{\mathrel{\hbox{\rlap{\hbox{\lower4pt\hbox{$\sim$}}}\hbox{$<$}}}}
\def\gtrsim{\mathrel{\hbox{\rlap{\hbox{\lower4pt\hbox{$\sim$}}}\hbox{$>$}}}}

\title{Gamma Ray Burst Central Engines}


\author{Todd A.~Thompson}{
  address={Department of Astronomy \& Center for Cosmology \& Astro-Particle Physics, \\
The Ohio State University \\
140 W.~18th Avenue, Columbus, Ohio 43210-1173\\
thompson@astronomy.ohio-state.edu}
}


%

\begin{abstract}
I review aspects of the theory of long-duration gamma-ray 
burst (GRB) central engines.  I focus on requirements of any model;
these include the angular momentum of the progenitor,
the power, Lorentz factor, asymmetry, and duration of the flow, 
and both the association and the non-association with bright 
supernovae.   I compare and contrast the collapsar and 
millisecond proto-magnetar models in light of these requirements.
The ability of the latter model to produce a flow with 
Lorentz factor $\sim$$100$ while simultaneously maintaining
a kinetic luminosity of $\sim10^{50}$\,ergs s$^{-1}$ for a 
timescale of $\sim10-100$\,s is emphasized.
\end{abstract}

\maketitle



\section{Introduction}
\label{section:intro}


Models for the central engine of long-duration gamma ray 
bursts (GRBs) are highly constrained by the character of  
the prompt emission and the afterglow, and --- at least in some cases
 --- the fact of an associated supernova (SN).  
There is little diversity among models for the central 
engine and essentially all can be simply classed as {\it a rotating 
compact object that drives an asymmetric relativistic outflow}. 

The ``collapsar'' model as described in \cite{Woosley93} and developed
in \cite{MW99,macfadyen_woosley_heger} proposes that GRBs arise from the collapse of rapidly 
rotating Type-Ibc progenitors. A black hole forms with an accompanying
accretion disk and drives a collimated relativistic outflow along the 
axis of rotation via either neutrino heating or magnetic stresses.

The ``millisecond proto-magnetar'' model 
posits a newly formed rotating
neutron star  (spin period $P\sim1$\,ms) 
with surface magnetic field of magnetar strength 
($B\sim10^{15}$\,G), cooling via neutrino radiation on the 
Kelvin-Helmholtz timescale, $t_{\rm KH}\sim10-100$\,s, and driving
a neutrino-heated magneto-centrifugal wind \cite{thompson04}.
Proto-magnetars might be produced by rotating Type-Ibc
progenitors, the accretion-induced collapse of a white dwarf,
the merger of two white dwarfs, and/or (potentially) the 
merger of two neutron stars \cite{MQT07}. Thus, 
they may trace both young and old stellar populations.
See also \cite{usov,thompson94,wheeler}.

Here, I compare and contrast some of 
the basic elements of and requirements on
the collapsar and proto-magnetar models.
I focus on the fact that a viable central engine must 
simultaneously realize a collimated flow with Lorentz factor 
in the range $100\lesssim \gamma\lesssim1000$, on a 
$10-100$ second timescale, while producing a
kinetic luminosity of $\sim10^{50}$\,ergs s$^{-1}$.
While many models have been proposed that can in
principle meet these requirements, 
the proto-magnetar model realizes them quantitatively.

\vspace{-.2cm}

\section{The Progenitor: Angular Momentum}

\vspace{-.2cm}

\paragraph{\bf Collapsar} The model
is that of a central black hole, formed by core-collapse of a massive
star, with a centrifugally-supported disk.  The latter requirement
sets the minimum angular momentum required for the collapsar
mechanism: $j_{\rm min}\simeq1.5\times10^{16}M_3$\,cm$^2$ s$^{-1}$
is required for disk formation at the ISCO of
a maximally-rotating $M_3=M/3$\,M$_\odot$ black hole.  For a disk that 
extends to larger radius and/or a non-maximal black hole, 
$j_{\rm min}$ increases.  The models of 
\cite{MW99} that produce a strong ``Ni wind,'' which
is needed to power the bright associated Type-Ibc SN 
lightcurve have $j\sim10^{17}$\,cm$^2$ s$^{-1}$.

\vspace{-.4cm}

\paragraph{\bf Proto-Magnetar}  The proto-magnetar model does
not require disk formation. Instead, it posits the existence
of a neutron star with a large-scale magnetic field of strength 
$B\sim10^{15}$\,G, mass $M_{1.4}=M/1.4$\,M$_\odot$, radius $R_{10}=R/10$\,km
and spin period of order $P_1=P/1$\,ms \cite{thompson04}.  
In a detailed set of calculations, \cite{M07a}
find that (depending on $B$) $P\lesssim2$\,ms 
is required to produce conditions favorable for a GRB,
impling a minimum specific angular momentum for 
the pre-collapse iron core of $j_{\rm min}\simeq3\times10^{15}
R_{10}^2P^{-1}_{1}$\,cm$^2$ s$^{-1}$.
This is a factor of 5 smaller than $j_{\rm min}$ for the collapsar model,
and a factor of $\sim30$ smaller than the model of \cite{MW99} 
that launches a disk wind.  
In absolute terms, a spin frequency ($\Omega$) 
correponding to $P\sim2$\,ms is relatively small, just $\sim25$\% of 
breakup, $\Omega_{\rm max}\sim(GM/R^3)^{1/2}$.

The additional requirement on the proto-magnetar model 
is that the neutron star must generate or be born with 
a magnetar-strength field. That large $B$ might accompany small $P$ 
was proposed by \cite{DT92,TD93}, who argued that when $P$
becomes shorter than the convective overturn timescale (Rossby number
$<1$) neutron stars undergo strong dynamo action, producing a magnetar.
A weakness of the proto-magnetar model is that a strong poloidal
field is assumed to exist without being generated 
self-consistently (e.g., \cite{thompson04,bucciantini06}).  
In absolute terms,  the energy associated with a $10^{15}$\,G field is much less than 
the rotational energy ($B_{{\rm eq,\,rot}}\sim [4\pi\rho R^2\Omega^2]^{1/2}
\sim2\times10^{17}\rho_{14}^{1/2}R_{10}P_{1}^{-1}\,\,{\rm G}$, 
where $\rho_{14}=\rho/10^{14}$\,g cm$^{-3}$) of the neutron star 
and/or the total energy carried in convective motions during $t_{\rm KH}$. 

\vspace{-.4cm}

\paragraph{\bf Summary} $j_{\rm min}$ for collapsars is significantly 
larger than for proto-magnetars.  If a progenitor stellar population
evolves in such a way as to produce conditions favorable for a 
collapsar, it is hard to see how it should not produce progenitors
with {\it smaller} specific angular momentum in the range needed
to power proto-magnetar-driven GRBs.  Perhaps in this way 
collapsars represent an extremum of the GRB population and 
proto-magnetars more continuously connect with the SN population.

\vspace{-.2cm}

\section{The Flow: Kinetic Power \& Lorentz Factor}

\vspace{-.2cm}

In the internal shock model, it is important that the flow 
that generates the GRB achieve $100\lesssim\gamma\lesssim1000$ while
{\it simultaneously} producing a kinetic 
luminosity of $\sim10^{50}$\,ergs s$^{-1}$ on a $t\sim10-100$\,s
timescale.  Reference  \cite{piran} writes that 
``A nagging question in all these models is what produces 
the 'observed' ultra-relativistic flow? How are 
$10^{-5}$\,M$_\odot$ of baryons accelerated to an 
ultra-relativistic velocity with  
$\gamma\sim100$ or larger? Why is the baryonic load 
so low? Why isn't it lower? There is no simple model for 
that. An ingenious theoretical idea is clearly needed here.''
See the lucid and concise discussion in \cite{nakar} of the physics that sets
the  upper and lower limits on $\gamma$.

\vspace{-.4cm}

\paragraph{\bf Collapsar}  There have been few
attempts to derive the Lorentz factor self-consistently
in the collapsar model.  Reference \cite{MW99} showed
that $\nu\bar{\nu}$ annihilation above the poles
of the black hole results in efficient energy deposition, 
generating very large 
specific entropy and high relativistic enthalpy.  It is 
natural that the flow should then accelerate 
to relativistic velocities, as in \cite{zhang03,morosony}.
A collimated relativistic flow might also be accelerated via 
magnetic stresses along the rotational axis, producing a 
highly Poynting-flux-dominated jet.  
Calculations of the global structure of 
accretion disks indicate that the polar funnel does realize
very large magnetization, $\sigma$, the magnitude of which is
limited only by numerics as the pole becomes essentially baryon-free
\cite{hawley_krolik}.
It has been suggested that cross-field neutron diffusion
might set the asymptotic Lorentz factor in an otherwise
baryon-free jet \cite{levinson_eichler};
however, see \cite{M07b}.

There is little doubt that the flow above the pole  of a rotating black 
hole will become highly relativistic as a result of either neutrino energy 
deposition or magnetic stresses, or both.  
The question is, 
does the collapsar model simultaneously produce 
a jet with $100\lesssim\gamma\lesssim1000$ and 
$\dot{E}\sim10^{50}$\,ergs s$^{-1}$?  What is 
the time-evolution of $\gamma$, and how does it relate 
to the time-dependence of the accretion rate?
These questions are as yet unanswered and they represent 
the largest single gap in the collapsar model of GRBs.

\vspace{-.4cm}

\paragraph{\bf Proto-Magnetar}
Regardless of the mechanism of core-collapse SNe, neutron stars are born
hot --- with a central temperature of 10's of MeV --- and they radiate their gravitational
binding energy in neutrinos as they cool, contract, and deleptonize on the 
Kelvin-Helmholtz timescale of $t_{\rm KH}\sim10-100$\,s \cite{burrows_lattimer,pons}.
During this epoch, neutron stars drive thermal neutrino-heated hydrodynamical winds 
\cite{duncan_shapiro_wasserman,QW96}.  For a neutron star with $M=1.4$\,M$_\odot$,
$R=10$\,km, rotation frequency $\Omega$, and 
and a magnetar-strength magnetic field, the mass-loss rate 
during $t_{\rm KH}$ is given approximately by \cite{QW96,thompson04,M07a}
\beq
\dot{M}(t)\approx10^{-6}\,L_{\bar{\nu}_e,\,51}^{5/2}(t)\,\,
\exp[\Omega(t)^2/\Omega_{\rm c}^2]\,\,\,{\rm M_\odot \,\,\,s^{-1}},
\label{mdot}
\eeq
where the total neutrino luminosity ($L_\nu$) is indexed by 
the $\bar{\nu}_e$ luminosity 
$L_{\bar{\nu}_e,\,51}=L_{\bar{\nu}_e}/10^{51}$\,ergs s$^{-1}$ 
(typically $L_\nu\sim5L_{\bar{\nu}_e}$) and it has
been assumed that $L_{\bar{\nu}_e}\propto\langle \varepsilon_{\bar{\nu}_e}\rangle^4$,
where $\langle \varepsilon_{\bar{\nu}_e}\rangle$ is the average
energy.  The normalization of 
eq.~(\ref{mdot}) and its $L_\nu$-dependence follow
from the physics of the weak interaction, and, in particular, the cross-section for 
the charged-current processes $\nu_e n\leftrightarrow p e^-$ and 
$\bar{\nu}_e p\leftrightarrow n e^+$ 
\cite{duncan_shapiro_wasserman,QW96}.  
This physics is a significant part of the answer to the questions posed
by \cite{piran} at the beginning of this section in the proto-magnetar model.
The exponential  factor in equation (\ref{mdot}) accounts for the enhancement by 
magneto-centrifugal forces when $B$ and $\Omega$ are large
($\Omega_{\rm c}\approx2300\,L_{\bar{\nu}_e,\,51}^{0.08}$\,rad s$^{-1}$;
\cite{M07a}).  For $P\lesssim2$\,ms
this factor becomes important \cite{thompson04}.

One of the most important components of the proto-magnetar model
is that as the neutron star cools $L_\nu$ decreases on a timescale $t_{\rm KH}$
\cite{burrows_lattimer,pons,thompson94,thompson04}.
As a result of equation (\ref{mdot}), $\dot{M}$ decreases concomitantly.
For this reason, for fixed surface magnetic field strength,  the wind
becomes increasingly magnetically-dominated and relativistic as a function of time.
This is quantified by the magnetization at the light
cylinder ($R_L=c/\Omega\sim50P_{1}$\,km): 
\beq
\sigma_L(t)=B(R_L)^2/[4\pi\rho(R_L) c^2]\propto \dot{M}(t)^{-1}.
\label{sigma}
\eeq 
At early times of order $\sim1$\,s after the SN shock is launched
$L_\nu$ is high enough that the wind is primarily 
driven by neutrino heating, $\dot{M}$ is large, $\sigma_L$ is less than unity, and the
flow is non-relativistic.  As $L_\nu$ decreases, the wind becomes
increasingly magnetically-dominated and it transitions to 
non-relativistic, but magneto-centrifugally dominated.  On a 
few-second timescale, $\sigma_L$ becomes larger than unity
and the wind becomes Poynting-flux dominated.  It becomes 
increasingly so on a timescale $t_{\rm KH}$. If we assume
that there is efficient conversion of the magnetic energy to 
bulk kinetic energy (via, e.g., magnetic dissipation; 
see \cite{drenkhahn_spruit}), then the asymptotic Lorentz 
factor of the wind is $\gamma\sim\sigma_L$.

As an example, in the models of \cite{M07a} (see, e.g., Fig.~1 of \cite{bucciantini07}),
we find that $\sigma_L\approx100$ at $t\approx20$\,s, when 
$\dot{E}\approx10^{50}$\,ergs s$^{-1}$,
for a proto-magnetar with $B=3\times10^{15}$\,G and $P=1$\,ms.
{\it Thus, proto-magnetars drive relativistic flows with the 
kinetic luminosity, timescale, and Lorentz factor appropriate for 
producing GRBs.}

The time evolution of the system is governed by perhaps the single most significant 
input to the proto-magnetar model: the time evolution of $L_\nu$ and 
$\langle \varepsilon_\nu\rangle$:  as $L_\nu$ decreases,  
$\dot{M}$ decreases (eq.~\ref{mdot}), $\sigma_L$ increases 
(eq.~\ref{sigma}), and, because $\dot{E}$ scales with a postive 
power of $\dot{M}$ \cite{thompson04,bucciantini06,M07a}, it  decreases. 
The cooling epoch eventually ends, $L_\nu$ and $\dot{M}$ drop 
precipitously, and the proto-magnetar transitions to more
``pulsar''-like.  As recently shown by \cite{M07b}, the fact 
that $\sigma_L(t)\sim\gamma(t)$ {\it increases} monotonically with time 
throughout $t_{\rm KH}$ implies high radiative efficiency in the 
internal shock model for the prompt emission.

In \cite{thompson04,M07a,M07b} we have taken $L_\nu(t)$ and $\langle \varepsilon_\nu\rangle(t)$
from \cite{pons}, informed by the early-time calculations of, e.g., \cite{tqb}.
However, the models of \cite{pons} are for {\it non-rotating non-magnetic} 
proto-neutron stars and it is possible that models including these effects 
(more appropriate for proto-magnetars) 
will change $L_\nu(t)$ and $\langle \varepsilon_\nu\rangle(t)$  quantitatively.

\vspace{-0.4cm}

\paragraph{\bf Summary} The ability of the proto-magnetar model to 
link $L_\nu$, $\dot{M}$, $\dot{E}$, and $\sigma_L\sim\gamma$
and their mutual time-dependence is a significant strength of the 
model and may allow it to be ruled out or (potentially) 
confirmed for some subsets of the global GRB population.

\vspace{-.4cm}

\section{The Explosion: (Non-)Association with Supernovae}

\vspace{-.2cm}

Several events link GRBs and SNe directly
(e.g., GRB 030329 \& SN 2003dh; \cite{stanek03}).  
Conversely, there are cases for which firm 
limits on an associated SN ($M[^{56}{\rm Ni}]<10^{-3}-10^{-4}$M$_\odot$)
can be established  (e.g., 060505, 060614; \cite{fynbo,galyam,gehrels}).

\vspace{-.4cm}

\paragraph{\bf Collapsar}  The collapsar model provides an
explanation for both the association and non-association with
SNe. High angular momentum cores produce extended collapsar disks, 
which launch non-relativistic winds as in \cite{MW99}. Calculations 
taking into account the electron fraction ($Y_e$) 
changing charged-current interactions show that even though
the midplane of the disk is neutron-rich, the outflow 
becomes de-neutronized with $Y_e\approx0.5$,
which then produces the $^{56}$Ni needed to power a SN lightcurve.
In contrast, low angular momentum cores  
that do not produce extended disks might not drive
Ni winds, although a relativistic jet above the poles 
is potentially possible.  Both scenarios
require a young stellar population.  This is qualitatively 
different from the proto-magnetar model.

\vspace{-.4cm}

\paragraph{\bf Proto-Magnetar}  
Some proto-magnetars may be formed from the collapse of a
Type-Ibc progenitor.  In this scenario, one imagines 
that the SN mechanism in some form (e.g., the ``neutrino mechanism''
\cite{bethe_wilson}) launches a SN shock with $\sim10^{51}$\,ergs
that produces $^{56}$Ni via explosive nucleosynthesis.  This would
explain the presence of an accompanying SN, but not the bright
character of some GRB-SNe.  If  $^{56}$Ni 
in excess of that produced in non-GRB Type-Ibc SNe is indeed
required in some cases (see \cite{soderberg_SN}) 
there are two ways in which it might be 
produced by proto-magnetars: (1) some of 
the initial rotational energy of the core may be tapped rapidly
via magnetic stresses, enhancing the 
SN shock energy as it is launched \cite{tqb,burrowsB} 
and/or (2) as the initial slow SN shockwave is moving outward it is 
shocked by the subsequent, highly-energetic proto-magnetar 
wind, again enhancing the shock energy \cite{thompson04}.  
Depending on the angular distribution of the wind kinetic energy, 
the latter option requires that $\gtrsim10^{51}$\,ergs 
is extracted from the proto-magnetar on a $\lesssim1-2$\,s timescale.

Proto-magnetars (and their GRBs) 
may also be formed by the accretion-induced collapse of white 
dwarfs and the merger of white dwarfs.  In this scenario 
there is no explosive nucleosynthesis, essentially no 
$^{56}$Ni yield, and no accompanying SN \cite{M07a,M07b}.  
Proto-magnetar GRBs could then potentially trace both
relatively young and old stellar populations.  This
is qualitatively different from the collapsar model.

\vspace{-0.4cm}

\paragraph{\bf Summary} 
The $^{56}$Ni yields of non-GRB Type-Ibc SNe are 
presumably generated by explosive nucleosynthesis.  
It would be remarkable if an entirely different 
mechanism --- i.e., the Ni wind from a collapsar disk ---  
was responsible for  $^{56}$Ni production in 
GRB-SNe given the quantitative similarlity between
the $^{56}$Ni yield distributions of GRB-SNe and non-GRB-SNe
\cite{soderberg_SN}.   Yet, this possibility is not excluded.
A potentially more natural explanation is that 
the similarity in $^{56}$Ni yields between 
GRB-SNe and non-GRB-SNe evidences the same underlying 
physical mechanism.  The proto-magnetar model provides
a simple and natural way of understanding this, as well as 
the association and the non-association of SNe with some GRBs.
It further allows for a simple interpretation of the 
continuum in properties between the explosive events 
of non-GRB-SNe and the GRB population itself 
(see, e.g., Fig.~5 of \cite{soderbergV}): the diversity 
in $B$ and $\Omega$ for neutron stars at birth leads to a
diversity in the amount of matter accelerated to relativistic
velocities and the energy and distribution of energy as a 
function of $\gamma$ for the explosion as a whole.

\vspace{-0.4cm}

\section{Asymmetry: Beaming \& Collimation}

\vspace{-0.2cm}

\paragraph{\bf Collapsar} Because of the geometry of the
system, the collapsar model provides a natural explanation 
for how the flow becomes highly-collimated, whether driven
by neutrino energy deposition \cite{MW99} or magnetic 
stresses \cite{hawley_krolik}.

\vspace{-0.4cm}

\paragraph{\bf Proto-Magnetar} The kinetic luminosity of a 
relativistic Poynting-flux-dominated proto-magnetar wind emerging into
vacuum is distributed broadly around the {\it equatorial plane} 
\cite{bucciantini06}; production of a jet is not trivial.
However, work by  \cite{bucciantini07,bucciantini08} shows that the 
interaction between the outflow and both the overlying stellar 
progenitor and the preceding SN shock acts to tightly
collimate the relativistic flow.  The mechanism was suggested in the GRB
context by \cite{konigl_granot}, based on models of  
\cite{begelman_li92}.
At radii much larger than $R_L$, the toroidal component of 
the wind magnetic field ($B_\phi$) is much larger than its polodal component.
The wind shocks on the exploding stellar envelope and produces a 
relativistically hot, quasi-hydrostatic bubble.  If $B_\phi$ is 
large enough, the bubble expands in the polar direction as a result
of both hoop stress at the equator and the confinement of the bubble
by the overlying stellar envelope.  The result is a relativistic jet.
The models of 
\cite{bucciantini08} indicate that little of the incident 
wind energy is coupled to the ``spherical''  component of the explosion (the SN) 
and, therefore, the asymptotic Lorentz factor of the jet reflects the 
Lorentz factor of the incident proto-magnetar wind.

\vspace{-0.4cm}

\paragraph{\bf Summary}
Based on analogy with observed jets in AGN and X-ray binaries
it may be argued that the picture of collimation in the 
collapsar scenario is more natural than that for the proto-magnetar.
Nevertheless,  \cite{bucciantini07,bucciantini08}  provide a 
compelling picture of proto-magnetar wind collimation.  
A major focus of future work on the proto-magnetar model
will be to understand the interaction of the wind on the progenitor
and the potential feedback on the wind itself if, for some 
parameters, the reverse shock propagates inside the fast,
Alfv\'en, and slow magnetosonic surfaces \cite{bucciantini08}.


\vspace{-0.4cm}

\begin{theacknowledgments}

\vspace{-0.3cm}

I thank B.~Metzger, N.~Bucciantini, E.~Quataert,
J.~Arons, and P.~Chang for collaboration on this 
project. This proceedings was motivated
in part by slides from S.~Woosley's
talk on GRB central engines at {\it Supernova 1987A:
20 Years After --- Supernovae and GRBs}.\footnote{
See http://astrophysics.gsfc.nasa.gov/conferences/supernova1987a/.}
\end{theacknowledgments}








\end{document}